\documentclass[12pt]{article} 
\usepackage[utf8]{inputenc} 
\usepackage[T1]{fontenc} 
\usepackage{amsmath,amssymb,bbm,bm,commath,mathtools,wasysym,xfrac} 
\usepackage{xcolor} 
\usepackage{graphicx} 
\graphicspath{{figures/}}
\usepackage{trimclip} 
\usepackage{stackengine} 
\usepackage{cite} 

\usepackage{sectsty} 
\usepackage{geometry} 
\usepackage{fancyhdr} 
\usepackage{setspace} 
\usepackage[nottoc,notlof,notlot]{tocbibind} 
\usepackage[titles]{tocloft} 
\usepackage[unicode]{hyperref} 
\hypersetup{bookmarksnumbered=true, bookmarksopen=true, bookmarksopenlevel=2, breaklinks=true, citecolor=blue, colorlinks=true, hypertexnames=false, linkcolor=red, linktoc=page, pdfborder={0 0 0}, pdfstartview=FitH, pdfauthor={DJain}, plainpages=false, unicode=true, urlcolor=cyan}

\DeclareUnicodeCharacter{0393}{\Gamma}
\DeclareUnicodeCharacter{0394}{\Delta}
\DeclareUnicodeCharacter{0398}{\Theta}
\DeclareUnicodeCharacter{039B}{\Lambda}
\DeclareUnicodeCharacter{039E}{\Xi}
\DeclareUnicodeCharacter{03A0}{\Pi}
\DeclareUnicodeCharacter{03A3}{\Sigma}
\DeclareUnicodeCharacter{03A5}{\Upsilon}
\DeclareUnicodeCharacter{03A6}{\Phi}
\DeclareUnicodeCharacter{03A8}{\Psi}
\DeclareUnicodeCharacter{03A9}{\Omega}
\DeclareUnicodeCharacter{03B1}{\alpha}
\DeclareUnicodeCharacter{03B2}{\beta}
\DeclareUnicodeCharacter{03B3}{\gamma}
\DeclareUnicodeCharacter{03B4}{\delta}
\DeclareUnicodeCharacter{03B5}{\epsilon}
\DeclareUnicodeCharacter{03B6}{\zeta}
\DeclareUnicodeCharacter{03B7}{\eta}
\DeclareUnicodeCharacter{03B8}{\theta}
\DeclareUnicodeCharacter{03D1}{\vartheta}
\DeclareUnicodeCharacter{03B9}{\iota}
\DeclareUnicodeCharacter{03BA}{\kappa}
\DeclareUnicodeCharacter{03BB}{\lambda}
\DeclareUnicodeCharacter{03BC}{\mu}
\DeclareUnicodeCharacter{03BD}{\nu}
\DeclareUnicodeCharacter{03BE}{\xi}
\DeclareUnicodeCharacter{03C0}{\pi}
\DeclareUnicodeCharacter{03C1}{\rho}
\DeclareUnicodeCharacter{03C3}{\sigma} 
\DeclareUnicodeCharacter{03C4}{\tau}
\DeclareUnicodeCharacter{03C5}{\upsilon}
\DeclareUnicodeCharacter{03C6}{\phi}
\DeclareUnicodeCharacter{03D5}{\varphi}
\DeclareUnicodeCharacter{03C7}{\chi}
\DeclareUnicodeCharacter{03C8}{\psi}
\DeclareUnicodeCharacter{03C9}{\omega}
\DeclareUnicodeCharacter{00B0}{^{\circ}}
\DeclareUnicodeCharacter{00B1}{\pm}
\DeclareUnicodeCharacter{00B7}{\cdot}
\DeclareUnicodeCharacter{00BD}{\tfrac{1}{2}}
\DeclareUnicodeCharacter{00D7}{\times}
\DeclareUnicodeCharacter{2020}{\dagger}
\DeclareUnicodeCharacter{2190}{\leftarrow}
\DeclareUnicodeCharacter{2192}{\rightarrow}
\DeclareUnicodeCharacter{2194}{\leftrightarrow}
\DeclareUnicodeCharacter{21D0}{\Leftarrow}
\DeclareUnicodeCharacter{21D2}{\Rightarrow}
\DeclareUnicodeCharacter{2202}{\partial}
\DeclareUnicodeCharacter{2208}{\in}
\DeclareUnicodeCharacter{220F}{\prod}
\DeclareUnicodeCharacter{222B}{\int}
\DeclareUnicodeCharacter{222E}{\oint}
\DeclareUnicodeCharacter{2211}{\sum}
\DeclareUnicodeCharacter{2213}{\mp}
\DeclareUnicodeCharacter{221E}{\infty}
\DeclareUnicodeCharacter{2243}{\simeq}
\DeclareUnicodeCharacter{2248}{\approx}
\DeclareUnicodeCharacter{2260}{\neq}
\DeclareUnicodeCharacter{2261}{\equiv}
\DeclareUnicodeCharacter{2264}{\leq}
\DeclareUnicodeCharacter{2265}{\geq}
\DeclareUnicodeCharacter{226A}{\ll}
\DeclareUnicodeCharacter{226B}{\gg}
\DeclareUnicodeCharacter{2297}{\otimes}
\DeclareUnicodeCharacter{22EF}{\cdots}
\DeclareUnicodeCharacter{25A1}{\square}
\DeclareUnicodeCharacter{266A}{\eighthnote}
\DeclareUnicodeCharacter{266B}{\twonotes}
\DeclareUnicodeCharacter{0212B}{\AA}

\definecolor{title}{rgb}{0.2,0.5,0.9}
\definecolor{abst}{rgb}{0.366,0.366,0.366}
\definecolor{sect}{rgb}{0.2,0.4,0.7}
\definecolor{ssect}{rgb}{0.4,0.5,1.0}
\definecolor{sssect}{rgb}{0.4,0.6,0.9}
\definecolor{appsect}{rgb}{0.9,0.2,0.5}
\definecolor{ref}{rgb}{0.0,0.0,1.0}
\definecolor{orcidlogocol}{HTML}{A6CE39}
\definecolor{azureblue}{HTML}{0066CC}
\definecolor{goldenyellow}{HTML}{FFCC00}
\newcommand{\Title}[1] {\title{\color{title}\Huge #1}}
\newcommand{\TPheader}[3] {\date{}\maketitle\thispagestyle{fancy}\pagenumbering{alph}\lhead{#1}\chead{#2}\rhead{#3}\cfoot{}}

\newcommand{\Abstract}[1] {\begin{abstract}\normalsize #1 \end{abstract}}
\newcommand{\makepage}[1] {\newpage\pagenumbering{#1}}

\sectionfont{\color{sect}}
\subsectionfont{\color{ssect}}
\subsubsectionfont{\color{sssect}}
\renewcommand{\appendix}{\setcounter{section}{0}\sectionfont{\color{appsect}}\renewcommand{\thesection}{\Alph{section}}\renewcommand*{\theHsection}{app.\the\value{section}}} 
\newcommand\references[1]{\sectionfont{\color{ref}}\bibliographystyle{hephys}\bibliography{#1}}

\newcommand\eqs[1] {\begin{align}#1\end{align}}

\newcommand\eqst[1] {\begin{multline}#1\end{multline}}

\newcommand\eqsa[1] {\equ{\begin{aligned}#1\end{aligned}}}

\newcommand\equ[1] {\begin{equation}#1\end{equation}}

\newcommand\fig[2] {\begin{figure}[#1]\centering #2\end{figure}}

\newcommand\s {\sigma}

\renewcommand\( {\left(}
\renewcommand\) {\right)}

\DeclareMathOperator{\Tr}{Tr}

\DeclareMathOperator{\tr}{tr}

\newcommand\bR {{\mathbb R}}

\newcommand\F {{\mathcal F}}

\renewcommand\O {{\mathcal O}}

\newcommand\W {{\mathcal W}}

\newcommand\fg {{\mathfrak g}}
\newcommand\fm {{\mathfrak m}}
\newcommand\fn {\mathfrak{n}}

\newcommand{\tu}{{\tilde{u}}}
\newcommand{\tnu}{{\tilde{\nu}}}
\newcommand\nn {\nonumber\\}

\newcommand\Hline {\noindent\rule{\textwidth}{0.5pt}}
\newcommand\forclr[7]{%
  \mbox{%
    \def\Whalf##1{\dimexpr.5\width ##1 #6em\relax}
    \def\Hhalf##1{\dimexpr.5\totalheight ##1 #7em\relax}
    \stackanchor[0pt]{\textcolor{#1}{\clipbox{0em {\Hhalf-} {\Whalf+} -0.1em}{$#5$}}%
    \textcolor{#2}{\clipbox{{\Whalf-} {\Hhalf-} 0em -0.1em}{$#5$}}}%
    {\textcolor{#4}{\clipbox{0em -0.1em {\Whalf+} {\Hhalf+}}{$#5$}}%
    \textcolor{#3}{\clipbox{{\Whalf-} -0.1em 0em {\Hhalf+}}{$#5$}}}%
  }%
}

\numberwithin{equation}{section} 
\begin{document}
\Title{Notes on 5d Partition Functions -- II}

\author{\href{mailto:dharmesh.jain@outlook.com}{Dharmesh Jain}\,\href{https://orcid.org/0000-0002-9310-7012}{\includegraphics[scale=0.0775]{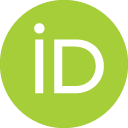}}\\
\emph{\normalsize Department of Theoretical Sciences, S. N. Bose National Centre for Basic Sciences,}\\
\emph{\normalsize Block--JD, Sector--III, Salt Lake City, Kolkata 700106, India}\bigskip\\
}

\TPheader{}{\texttt{1--04--2022}}{} 

\Abstract{We study the large $N$ limit of partition functions for 5d supersymmetric gauge theories with fundamental matter. Depending on the matter content, we find that the scaling behaviour at the leading order can be either $N^2$ or $N^{\frac{3}{2}}$. The latter scaling reminds one of the 3d theories with M-theory duals and we discuss how to extract this behaviour from the recently proposed 3d theories associated with the compactification of 5d SCFTs on 2d surfaces.\\

\hfill \emph{Dedicated to\forclr{azureblue}{azureblue}{goldenyellow}{goldenyellow}{PEACE}{0}{0}}
}

\Hline\vspace*{-6mm}\tableofcontents\Hline 
\makepage{arabic} 

\section{Introduction and Summary}
The five-dimensional supersymmetric gauge theories possess interesting dynamics. Even though they are power-counting nonrenormalizable and are IR trivial in general, some 5d gauge theories are known to arise as a relevant deformation of nontrivial UV 5d SCFTs, while some are even known to be UV completed by 6d SCFTs. Furthermore, there is holographic evidence for the existence of RG flow across dimensions from 5d theories to 3d SCFTs. This duality across dimensions or ``5d/3d correspondence'' has recently become quite an interesting playground to study certain classes of 5d theories. The holographic prediction was verified at large $N$ for certain classes of theories, including Seiberg theories \cite{Seiberg:1996bd} and related quiver theories \cite{Bergman:2012kr}, by explicit field theoretical computation of $S^3_b×Σ_{\fg}$ partition function in \cite{Crichigno:2018adf}. Similar results for different 5d manifolds have been obtained in the large $N$ limit \cite{Hosseini:2018uzp,Jain:2021sdp} and in Cardy limit \cite{Hosseini:2021mnn}.

All the above-mentioned results inherently study and compute 5d partition functions without dealing directly with the underlying 3d theory (if any). This issue is tackled head-on in \cite{Sacchi:2021afk,Sacchi:2021wvg} with explicit construction of the 3d theories whose global symmetries, operator spectra, etc. are expected to match the corresponding 5d gauge theories compactified on 2d surfaces (specifically, tubes and tori). The 5d theories considered in these articles do not belong to the classes of theories mentioned above whose 5d partition functions are easily evaluated. The difficulty arises because of insufficient cancellation of nonlocal terms in the matrix models that emerge in the large $N$ limit of the partition functions. However, since explicit 3d theories are conjectured to exist, it seems like a worthwhile exercise to compute partition functions for both 5d and 3d theories and check the correspondence. In this short note, we will focus on the first half of the correspondence and compute $S^5_{\vec{ω}}$ and $S^3_{b}×Σ_{\fg}$ partition functions of a few simple 5d theories; those discussed extensively in \cite{Intriligator:1997pq}. We find that the large $N$ scaling of these theories behaves as $N^2$ or $N^{\frac{3}{2}}$, depending on the number of fundamental matter multiplets, unlike the more familiar $N^{\frac{5}{2}}$ scaling of various quiver theories with $AdS_6$ duals. We will comment on the other half of the correspondence involving $S^3_b$ partition function of 3d theories at the end of this note but leave its detailed analysis to \cite{Jain:2022td}.

\paragraph{Outline.} This note is organized as follows. In Section \ref{sec:Review} we review the relevant 5d theories, whose $S^5_{\vec{ω}}$ and $S^3_{b}×Σ_{\fg}$ partition functions are then computed in Section \ref{sec:S5pf} and \ref{sec:S3Spf}, respectively. In Section \ref{sec:S3pf}, we then discuss a sample computation of $S^3_b$ partition function of a conjectured 3d theory associated with the compactification of a specific 5d SCFT.

\section{Review}\label{sec:Review}
We review some aspects of the relevant 5d supersymmetric gauge theories following \cite{Intriligator:1997pq}. The 5d exact low-energy effective prepotential for gauge group $G$ and matter hypermultiplets (HM) in representation $R_I$ with masses $m_I$ is given by
\equ{\F =\frac{1}{2g^2}\Tr(\s^2) +\frac{k}{6}\Tr(\s^3) +\frac{1}{12}\bigg(∑_{α∈Ad(G)'}|α·\s|^3 -∑_I∑_{ρ∈R_I}|ρ·\s+m_I|^3\bigg),
}
where $\s$ is the (adjoint) scalar in vector multiplet (VM), $α$ are the roots of $G$, $ρ$ are the weights of $G$ in rep $R_I$. The first two terms arise from the 5d supersymmetric Yang-Mills (YM) and Chern-Simons (CS) action. The last two terms are the one-loop quantum contributions. It was shown in \cite{Intriligator:1997pq} that the existence of a nontrivial UV fixed point for such theories requires that the 5d prepotential should be a convex function over the entire Coulomb branch (Weyl chamber). The convexity of the 5d prepotential is guaranteed when its Hessian $\big(\frac{∂^2\F}{∂\s_i∂\s_j}\big)$ has non-negative eigenvalues. This analysis leads to constraints on the possible matter content for various gauge groups along with $g^{-2}=0$. These constraints can be relaxed slightly as discussed in \cite{Bergman:2014kza,Hayashi:2015fsa,Bergman:2015dpa}. However, instanton contributions play a crucial role in such relaxations which we will not discuss in this note. We restrict ourselves to the results from \cite{Intriligator:1997pq}, which are summarized below:
\begin{description}
\item[$\bm{SU(N)}$.] The Coulomb branch is given by $\s=\text{diag}(\s_1,⋯,\s_N)$ with $∑_i\s_i=0$, modulo permutations, which is the Weyl group action. Thus, we can take the Weyl chamber to be $\s_1≥\s_2≥⋯≥\s_N$. Considering theories with $N_s$ symmetric HMs, $N_{as}$ antisymmetric HMs and $N_f$ fundamental HMs, the Hessian is calculated easily. Along the direction where $\s_i=λ$, $i=1,⋯,N-1$, such that $SU(N)$ is broken to $SU(N-1)×U(1)$, the non-negativity of the eigenvalues of the Hessian produces the following constraints:
\equ{N_s=0\,,\quad N_{as}=0\,,\quad N_f≤2N-2|k|\,.
\label{SUNconst}}
This is the only constraint that is applicable when $N$ is large. There are constraints leading to $N_{as}≠0$ when $N≤8$, but we will not be interested in these `finite $N$' theories. The absolute value on $k$ takes care of the charge conjugation operation, which transforms $\s_i→-\s_{N+1-i}$ and $k→-k$. For the rest of the note, we consider $k$ to be positive, unless otherwise stated, without loss of generality. 

\item[$\bm{USp(2N)}$.] The Coulomb branch is given by $\s=\text{diag}(\s_1,⋯,\s_N,-\s_1,⋯,-\s_N)$, modulo the Weyl group action. Thus, we can take the Weyl chamber to be $\s_1≥\s_2≥⋯≥\s_N≥0$ and consider theories with $N_{as}$ antisymmetric HMs and $N_f$ fundamental HMs. Along the direction where $\s_i=λ$ for $i=1,⋯,p$ and $\s_{i>p}=0$, $USp(2N)$ is broken to $USp(2(N-p))×SU(p)×U(1)$, the constraints turn out to be
\eqsa{N_{as}=0\,,& \quad N_f≤2N+4\,; \\
N_{as}=1\,, &\quad N_f≤7\,.
\label{USp2Nconst}}
The latter case has been studied in great detail starting from the work of \cite{Jafferis:2012iv} so we will only focus on the former case with no antisymmetric HMs.

\item[$\bm{SO(M)}$.] We take $M$ to be of the form $2N+δ$ with $δ$ being 0(1) for even(odd) $M$. The Coulomb branch is given by the Weyl chamber $\s_1≥\s_2≥⋯≥\s_N≥0$ and we consider theories with $N_f$ fundamental (vector) HMs and $N_{sp}$ spinorial HMs. Along the direction where $\s_i=λ$ for $i=1,⋯,p$ and $\s_{i>p}=0$, $SO(M)$ is broken to $SO(M-2p)×U(p)$, the constraints turn out to be
\eqsa{N_f≤M-4\,,\quad N_{sp}≤2^{6-N-δ}\,.
\label{SOMconst}}
Since we are interested in the large $N$ limit, it is clear that spinorial HMs are not allowed.
\end{description}
Having recalled the various possible 5d theories, we now move on to studying their 5d partition functions.

\section{\texorpdfstring{$\bm{S^5_{\vec{ω}}}$ Free Energy}{S⁵ω Free Energy}}\label{sec:S5pf}
The construction of the large $N$ expression for free energy on $S^5$ follows from \cite{Jafferis:2012iv,Imamura:2013xna,Alday:2014bta}. Let us begin with the definition of free energy $F$:
\begingroup
\allowdisplaybreaks
\eqs{Z_{S^5_{\vec{ω}}} &=\frac{1}{|\W|}∫d\s^i e^{-F_{S^5_{\vec{ω}}}(\s)}\,, \nn
F_{S^5_{\vec{ω}}}(\s)&≡\frac{4π^2r}{g^2ω_1ω_2ω_3}\tr_F\s^2 +\frac{πk}{3ω_1ω_2ω_3}\tr_F\s^3 +\tr_{Ad}F_V(\s) +∑_I\tr_{R_I}F_H(\s)\,.
}
\endgroup
The localization computation has reduced the full path integral to just integrals over the scalar $\s$ in the Cartan of the gauge algebra. For large $N$, we only need the $F_V,F_H$ functions at large argument ($|\s|≫1$), which read as follows:
\eqsa{F_V(\s) &≈\frac{π}{6ω_1ω_2ω_3}|\s|^3 -\frac{(ω_{sum}^2+ω_{sym}^2)π}{12ω_1ω_2ω_3}|\s|\,, \\
F_H(\s) &≈-\frac{π}{6ω_1ω_2ω_3}|\s|^3 -\frac{(ω_{sum}^2-2ω_{sym}^2)π}{24ω_1ω_2ω_3}|\s|\,,
}
where we use the notation $ω_{sum}=ω_1+ω_2+ω_3$ and $ω_{sym}^2=ω_1ω_2+ω_2ω_3+ω_3ω_1$. With these building blocks, the free energy can be easily written down for the above-mentioned theories. However, evaluating / extremizing the resulting expressions using the continuum approach as discussed in \cite{Jafferis:2012iv,Uhlemann:2019ypp} does not seem to give sensible results. So we simplify the evaluation procedure drastically by equating some of the eigenvalues $\s_i$ and setting the rest to zero (as done in the analysis of \cite{Intriligator:1997pq} to obtain the constraints reviewed above), to obtain a free energy expression depending on only one variable, which can then be easily extremized.\footnote{This simplified approach also gives the correct $N^{\frac{5}{2}}$ scaling for the `usual' $USp(2N)$ theories satisfying the second set of constraint in \eqref{USp2Nconst}, but the coefficients do not match as one might expect. A somewhat different approach has also appeared in \cite{Minahan:2014hwa,Nedelin:2015mta,Santilli:2021qyt} for 5d theories with adjoint and fundamental matter, which exhibit $N^2$ scaling.} We now specialize to the three classes of 5d gauge theories discussed above.

\subsection[\texorpdfstring{$SU(N)_k+N_f$}{SU(N)k+Nf}]{$\bm{SU(N)_k+N_f}$}
Consider a $SU(N)$ Yang-Mills theory with Chern-Simons level $k$ (assumed to be $\O(1)$, unless otherwise stated) with $N_f$ fundamental hypermultiplets. The VM scalar is parametrized as $\s=\text{diag}\(\s_1,⋯,\s_N\)$ with $\s_N=-∑_{i=1}^{N-1}\s_i$ such that the $S^5$ free energy can be written down as follows (setting $g^{-2}=0$ from now on):
\equ{F^{SU}_{S^5_{\vec{ω}}}(\s) =∑_{i=1}^{N}\left[\frac{πk}{3ω_1ω_2ω_3}\s_i^3 +∑_{j=1}^{i-1}2F_V(\s_i-\s_j) +∑_{I=1}^{N_f}F_H(\s_i)\right].
\label{Fsu}}
We used that the adjoint rep of $SU(N)$ has roots $±(e^i-e^j)$ ($j<i$) and fundamental rep has weights simply given by $e^i$, with $\{e^i\}$ being the basis of unit vectors in $\bR^N$. Further restricting to the Weyl chamber $\s_1≥\s_2≥⋯≥\s_N$ and choosing $\s_i=λ$ for $i=1,⋯,N-1$ (as discussed above), we get
\eqst{F^{SU}_{S^5_{\vec{ω}}}(\s) =-\frac{πkN(N-1)(N-2)}{3ω_1ω_2ω_3}λ^3 +\(\frac{π(N-1)N^3}{3ω_1ω_2ω_3}λ^3 -\frac{π(ω_{sum}^2+ω_{sym}^2)N(N-1)}{6ω_1ω_2ω_3}λ\) \\
 -N_f\(\frac{π(N-1)(N^2-2N+2)}{6ω_1ω_2ω_3}λ^3 +\frac{π(ω_{sum}^2-2ω_{sym}^2)(N-1)}{12ω_1ω_2ω_3}λ\).
\label{Fsusimp}}
We can now extremize the above free energy wrt $λ$ easily and obtain the extremized free energy to be
\equ{\bar{F}^{SU}_{S^5_{\vec{ω}}} =-\frac{π(N-1)[2N(ω_{sum}^2+ω_{sym}^2)+N_f(ω_{sum}^2-2ω_{sym}^2)]^{\frac{3}{2}}}{18ω_1ω_2ω_3\sqrt{3}\sqrt{4N^3-4k N(N-2) -2N_f(N^2-2N+2)}}\,·
}
Demanding the denominator does not vanish reproduces the allowed values of $N_f$:
\equ{N_f≤2N+4-2k\,.
}
Comparing this to \eqref{SUNconst}, we see that the above constraint is less strict but matches the one found in \cite{Bergman:2014kza,Hayashi:2015fsa,Bergman:2015dpa}. Continuing to ignore instanton corrections and assuming $N_f$ to be of the form $n N+f$, we see the extremized free energy takes the following form:
\equ{\bar{F}^{SU}_{S^5_{\vec{ω}}} =-\frac{π(N-1)[N\{(2+n)ω_{sum}^2+2(1-n)ω_{sym}^2\}+f(ω_{sum}^2-2ω_{sym}^2)]^{\frac{3}{2}}}{18ω_1ω_2ω_3\sqrt{3}\sqrt{(4-2n)N^3 +2(2n-2k-f)N^2 -4(n-2k-f)N -4f}}\,·
}
This allows us to discuss the large $N$ limit of $\bar{F}_{S^5}$ for the sequence of theories considered in \cite{Sacchi:2021wvg} and this limit falls into three cases:
\begin{enumerate}
\item When $n=2$ with $f=4-2k$:
\equ{\bar{F}^{SU}_{S^5_{\vec{ω}}}=-\frac{π(2ω_{sum}^2-ω_{sym}^2)^{\frac{3}{2}}}{18\sqrt{3}\,ω_1ω_2ω_3}N^2\,.
}
In the range $2N-2k<N_f≤2N+4-2k$, the above result may not be the whole story owing to instanton contributions. Also, the UV completion is supposed to be a 6d SCFT in this range so $N^3$ scaling is more likely.
\item When $n=2$ with $f<4-2k$:\footnote{If $k$ happens to be $\O(N)$, we also get $N^{\frac{3}{2}}$ scaling.}
\equ{\bar{F}^{SU}_{S^5_{\vec{ω}}}=-\frac{π(2ω_{sum}^2-ω_{sym}^2)^{\frac{3}{2}}}{9\sqrt{3(4-2k-f)}\,ω_1ω_2ω_3}N^{\frac{3}{2}}\,.
}
\item When $0≤n<2$ and $f$ is $\O(1)$:
\equ{\bar{F}^{SU}_{S^5_{\vec{ω}}}=-\frac{π[(2+n)ω_{sum}^2+2(1-n)ω_{sym}^2]^{\frac{3}{2}}}{18\sqrt{3(4-2n)}\,ω_1ω_2ω_3}N\,.
}
This result is definitely not the whole story as instantons are expected to contribute at $\O(N)$.
\end{enumerate}

\subsection[\texorpdfstring{$USp(2N)+N_f$}{USp(2N)+Nf}]{$\bm{USp(2N)+N_f}$}
This is a $USp(2N)$ Yang-Mills theory with matter consisting of zero antisymmetric hypermultiplets but only $N_f$ fundamental hypermultiplets. The VM scalar is parametrized as $\s=\text{diag}\(\s_1,⋯,\s_N,-\s_1,⋯,-\s_N\)$ such that the free energy becomes (with $k=0$):
\equ{F^{USp}_{S^5_{\vec{ω}}}(\s) =∑_{±,i,j<i}\left[F_V(±(\s_i-\s_j)) +F_V(±(\s_i+\s_j))\right] +∑_{±,i}F_V(±2\s_i) +∑_{±,i}N_f F_H(±\s_i)\,,
\label{Fusp}}
where we used that the adjoint rep of $USp(2N)$ has roots $±(e^i±e^j)$ ($j<i$) and $±2e^i$, and fundamental rep has weights $±e^i$. Further restricting to the Weyl chamber $\s_1≥\s_2≥⋯≥\s_N≥0$ and choosing $\s_i=λ$ for $i=1,⋯,p$ with $\s_{i>p}=0$ (as discussed in the previous section), we get
\eqst{F^{USp}_{S^5_{\vec{ω}}}(\s) =\(\frac{2πp(N+p-2)}{3ω_1ω_2ω_3}λ^3 -\frac{πp(2N-p-1)(ω_{sum}^2+ω_{sym}^2)}{6ω_1ω_2ω_3}λ\) \\
+\(\frac{π(8-N_f)p}{3ω_1ω_2ω_3}λ^3 -\frac{πp[(4+N_f)ω_{sum}^2+(4-2N_f)ω_{sym}^2]}{12ω_1ω_2ω_3}λ\).
\label{Fuspsimp}}
Again, we can extremize the above free energy straightforwardly and obtain the extremized result as follows
\equ{\bar{F}^{USp}_{S^5_{\vec{ω}}} =-\frac{πp[(4N+2-2p+N_f)ω_{sum}^2+2(2N+1-p-N_f)ω_{sym}^2]^{\frac{3}{2}}}{36\sqrt{3(2N+4+2p-N_f)}\,ω_1ω_2ω_3}\,·
}
Demanding the denominator does not vanish constrains the possible values of $N_f$:
\equ{N_f≤2N+4+2p\,,
}
which is consistent with \eqref{USp2Nconst}. We can now discuss the large $N$ limit of $\bar{F}_{S^5}$ for various possible theories and this limit again falls into three cases:
\begin{enumerate}
\item When $N_f=n N+f$ with $0≤n≤2$ and $f$ being $\O(1)$ but $p=N-n_p$ with $0≤n_p≪N$:
\equ{\bar{F}^{USp}_{S^5_{\vec{ω}}}=-\frac{π[(2+n)ω_{sum}^2+2(1-n)ω_{sym}^2]^{\frac{3}{2}}}{36\sqrt{3(4-n)}\,ω_1ω_2ω_3}N^2\,.
}
A special case is $n=2$ and $f≤4$ which gives
\equ{\bar{F}^{USp}_{S^5_{\vec{ω}}}=-\frac{π(2ω_{sum}^2-ω_{sym}^2)^{\frac{3}{2}}}{18\sqrt{3}\,ω_1ω_2ω_3}N^2\,,
}
similar to the case 1. for $SU(N)_k+N_f$ theories.
\item When $N_f=2N+f$ with $f≤4$ but $p$ being $\O(1)$:
\equ{\bar{F}^{USp}_{S^5_{\vec{ω}}}=-\frac{πp\,ω_{sum}^3}{3\sqrt{2(4+2p-f)}\,ω_1ω_2ω_3}N^{\frac{3}{2}}\,.
}
\item When $N_f=n N+f$ with $0≤n<2$ and $f,p$ both being $\O(1)$:
\equ{\bar{F}^{USp}_{S^5_{\vec{ω}}}=-\frac{πp[(4+n)ω_{sum}^2+2(2-n)ω_{sym}^2]^{\frac{3}{2}}}{36\sqrt{3(2-n)}\,ω_1ω_2ω_3}N\,.
}
Once again, this result will get corrected with instantons contributions at this order.
\end{enumerate}

\subsection[\texorpdfstring{$SO(M)+N_f$}{SO(M)+Nv}]{$\bm{SO(M)+N_f}$}
This is a $SO(M)$ Yang-Mills theory with matter consisting of only $N_f$ vector hypermultiplets. We also use $M=2N+δ$ with $δ$ being 0(1) for even(odd) $M$. The VM scalar is then parametrized as $\s=\text{diag}\{\s_1,⋯,\s_N\}$ with the Weyl chamber chosen as $\s_1≥\s_2≥⋯≥\s_N≥0$. As discussed in the previous section, we choose $\s_i=λ$ for $i=1,⋯,p$ with $\s_{i>p}=0$. With this setup, the free energy becomes (again $k=0$)
\eqs{F^{SO}_{S^5_{\vec{ω}}}(\s) &=∑_{±,i,j<i}\left[\frac{π}{3ω_1ω_2ω_3}(\s_i±\s_j)^3 -\frac{π(ω_{sum}^2+ω_{sym}^2)}{6ω_1ω_2ω_3}(\s_i±\s_j)\right] \nn
&\quad+∑_i\(\frac{π(δ-N_f)}{3ω_1ω_2ω_3}\s_i^3 -\frac{π[(2δ+N_f)ω_{sum}^2+2(δ-N_f)ω_{sym}^2]}{12ω_1ω_2ω_3}\s_i\) \nn
&=\(\frac{2πp(N+p-2)}{3ω_1ω_2ω_3}λ^3 -\frac{πp(2N-p-1)(ω_{sum}^2+ω_{sym}^2)}{6ω_1ω_2ω_3}λ\) \nn
&\quad+\(\frac{πp(δ-N_f)π}{3ω_1ω_2ω_3}λ^3 -\frac{πp[(2δ+N_f)ω_{sum}^2+2(δ-N_f)ω_{sym}^2]}{12ω_1ω_2ω_3}λ\),
\label{Fspin}}
where we used that the adjoint rep of $SO(M)$ has roots $±(e^i±e^j)$ ($j<i$) and $±δe^i$, and vector rep has weights $±e^i$. Again, we can extremize the above free energy straightforwardly and obtain the extremized result as follows
\equ{\bar{F}^{SO}_{S^5_{\vec{ω}}} =-\frac{πp[(2(2N+δ)-2p-2+N_f)ω_{sum}^2 +(2N+δ-p-1-N_f)ω_{sym}^2]^{\frac{3}{2}}}{36\sqrt{3(2N+δ+2p-4-N_f)}\,ω_1ω_2ω_3}\,·
}
Demanding the denominator does not vanish constrains the possible values of $N_f$:
\equ{N_f≤M-4+2p\,,
}
which is consistent with \eqref{SOMconst}. We again find the large $N$ limit of $\bar{F}_{S^5}$ falls into three cases (very similar to the $USp(2N)$ case):
\begin{enumerate}
\item When $N_f=n N-f$ with $0≤n≤2$ and $f$ being $\O(1)$ but $p=N-n_p$ with $0≤n_p≪N$:
\equ{\bar{F}^{SO}_{S^5_{\vec{ω}}}=-\frac{π[(2+n)ω_{sum}^2+2(1-n)ω_{sym}^2]^{\frac{3}{2}}}{36\sqrt{3(4-n)}\,ω_1ω_2ω_3}N^2\,.
}
A special case is $n=2$ and $f≥4-δ$ which gives
\equ{\bar{F}^{SO}_{S^5_{\vec{ω}}}=-\frac{π(2ω_{sum}^2-ω_{sym}^2)^{\frac{3}{2}}}{18\sqrt{3}\,ω_1ω_2ω_3}N^2\,.
}
\item When $N_f=2N+δ-f$ with $f≥4-δ$ but $p$ being $\O(1)$:
\equ{\bar{F}^{SO}_{S^5_{\vec{ω}}}=-\frac{πp\,ω_{sum}^3}{3\sqrt{2(2p+f-4)}\,ω_1ω_2ω_3}N^{\frac{3}{2}}\,.
}
\item When $N_f=n N+f$ with $0≤n<2$ and $f,p$ both being $\O(1)$:
\equ{\bar{F}^{SO}_{S^5_{\vec{ω}}}=-\frac{πp[(4+n)ω_{sum}^2+2(2-n)ω_{sym}^2]^{\frac{3}{2}}}{36\sqrt{3(2-n)}\,ω_1ω_2ω_3}N\,.
}
This result at $\O(N)$ will get corrected with instanton contributions.
\end{enumerate}
As we clearly see, the $SO(M)$ results mirror those of $USp(2N)$ results in almost every detail so we will not discuss the $SO(M)$ theories in the upcoming sections.

\section{\texorpdfstring{$\bm{S^3_b×Σ_{\fg}}$ Free Energy}{S³b×Σg Free Energy}}\label{sec:S3Spf}
We collect some relevant results for the $S^3_b×Σ_{\fg}$ partition function from \cite{Crichigno:2018adf}:
\eqst{Z_{S^3_b×Σ_{\fg}} =\frac{1}{|\W|}∑_{\fm^i}∮d\tu^ie^{-\frac{4π^2}{g^2}Q^2\fm·\tu +iπkQ^2\Tr(\fm\tu^2)}∏_{α∈Ad(G)'}s_b\(-iQ(α(\tu)+1)\)^{1-\fg-α(\fm)} \\
×∏_I∏_{ρ∈R_I}s_b\(-iQ(ρ(\tu)+\tnu_I)\)^{ρ(\fm)+\fn_I(\fg-1)}\,,
}
where $\tu$ is the gauge variable, $\fm(\fn)$ is the gauge(flavour) magnetic flux and $\tnu$ is the flavour fugacity. For large $N$, we need the asymptotic behaviour of the $s_b(iQz)=e^{\ell_b(z)}$ function:
\equ{\ell_b(a+i λ) ≈∓\(\frac{iπ}{2}λ^2 +πaλ -\frac{iπ}{2}a^2 +\frac{iπ}{24}(b^2+b^{-2})\) \quad\text{ for }λ→±∞.
\label{eq4p2S2B22}}
With this setup, we now specialize to the 5d gauge theories discussed in previous sections.

\subsection[\texorpdfstring{$SU(N)_k+N_f$}{SU(N)k+Nf}]{$\bm{SU(N)_k+N_f}$}
For this theory, we write the $S^3_b×Σ_{\fg}$ free energy by taking the log of the partition function given above ($g^{-2}=0$ as before):
\eqst{F_{S^3_b×Σ_{\fg}}^{SU} =∑_{i=1}^N\(-iπkQ^2\fm_i\tu_i^2\) -∑_{i=1}^N∑_{j=1}^{i-1}(1-\fg∓\fm_i ±\fm_j)\ell_b\(1±(\tu_i-\tu_j)\) \\
+N_f∑_{i=1}^N (\fm_i+(\fg-1)\fn_f)\ell_b\(\tu_i+\tnu_f\).
}
We can follow the approach of extremizing the twisted superpotential first and then evaluating the free energy on those solutions but for the purposes of swiftness, we follow the approach of extremizing the free energy with respect to both the gauge variable $\tu$ and gauge flux $\fm$ following \cite{Hosseini:2021mnn}. From past experiences, we expect the extremum values for $\tu$ to be imaginary so we substitute $\tu_i→i\s_i$. Furthermore, we also choose the ansatz $\fm_i→iη\s_i$ and restrict to the Weyl chamber $\s_1≥\s_2≥⋯≥\s_N≥0$ along with $\s_i=λ$ for $i=1,⋯,N-1$ (as before). We now have to extremize wrt both $λ$ and $η$, which is a straightforward exercise after using \eqref{eq4p2S2B22}, and we get
\equ{\bar{F}_{S^3_b×Σ_{\fg}}^{SU} =(\fg-1)\frac{2πQ(N-1)(N-N_f\fn_f\tnu_f)\sqrt{2N(4Q^2+1)+N_f(4Q^2-2) -12Q^2N_f\tnu_f^2}}{\sqrt{3}\sqrt{4N^3-4k N(N-2) -2N_f(N^2-2N+2)}}\,·
}
Demanding the denominator does not vanish reproduces the same constraint on $N_f$ as before: $N_f≤2N+4-2k$. So we can discuss the large $N$ limit of $\bar{F}_{S^3×Σ_{\fg}}$ and compare with the $\bar{F}_{S^5}$ for the three cases as follows ($N_f=n N+f$):
\begin{enumerate}
\item When $n=2$ with $f=4-2k$:
\equ{\bar{F}_{S^3_b×Σ_{\fg}}^{SU} =(\fg-1)\frac{πQ}{\sqrt{3}}(1-2\fn_f\tnu_f)\sqrt{8Q^2-1-12Q^2\tnu_f^2}\,N^2\,.
}
\item When $n=2$ with $f<4-2k$:
\equ{\bar{F}_{S^3_b×Σ_{\fg}}^{SU} =(\fg-1)\frac{2πQ}{\sqrt{3}\sqrt{4-2k-f}}(1-2\fn_f\tnu_f)\sqrt{8Q^2-1-12Q^2\tnu_f^2}\,N^{\frac{3}{2}}\,.
}
\item When $0≤n<2$ with $f$ being $\O(1)$:
\equ{\bar{F}_{S^3_b×Σ_{\fg}}^{SU} =(\fg-1)\frac{2πQ}{\sqrt{3}\sqrt{4-2n}}(1-n\fn_f\tnu_f)\sqrt{4(2+n)Q^2+2(1-n)-12nQ^2\tnu_f^2}\,N \,.
}
As instantons are expected to contribute at $\O(N)$, we may not trust this result at this order.
\end{enumerate}
To compare with $\bar{F}_{S^5_{\vec{ω}}}^{SU}$ results, one can set $ω_1=i,ω_2=-i,ω_3=2Q$ $⇒ω_{sum}^2=4Q^2,ω_{sym}^2=1$ and identify the common factors. More concretely, one can construct the twisted superpotential $\W_{S^3_b×\bR^2}$ that turns out to be proportional to $\bar{F}_{S^5_{\vec{ω}}}$ and then one can relate the two free energies in the usual manner: $\bar{F}_{S^3_b×Σ_{\fg}} \propto \frac{∂\bar{\W}_{S^3_b×\bR^2}}{∂\tnu_f}\,·$ We leave this exercise to the reader.

\subsection[\texorpdfstring{$USp(2N)+N_f$}{USp(2N)+Nf}]{$\bm{USp(2N)+N_f}$}
Repeating the above analysis with previous sections' conventions for $USp(2N)$ theories, we get
\eqst{\bar{F}_{S^3_b×Σ_{\fg}}^{USp} =(\fg-1)\frac{πQp(2N+1-p-N_f\fn_f\tnu_f)}{\sqrt{3}\sqrt{2N+4+2p -N_f}}× \\
\sqrt{4(4N+2-2p+N_f)Q^2 +2(2N+1-p-N_f) -12Q^2N_f\tnu_f^2}\,.
}
This allows us to write down the large $N$ limit of $\bar{F}_{S^3×Σ_{\fg}}$ for the three cases as follows ($N_f=n N+f$):
\begin{enumerate}
\item When $0≤n≤2$ with $f$ being $\O(1)$ but $p=N-f_p$ with $0≤f_p≪N$:
\equ{\bar{F}^{USp}_{S^3×Σ_{\fg}}=(\fg-1)\frac{πQ(1-n\fn_f\tnu_f)}{\sqrt{3}\sqrt{4-n}}\sqrt{4(2+n)Q^2 +2(1-n) -12nQ^2\tnu_f^2}\,N^2\,.
}
\item When $n=2$ with $f≤4$ but $p$ being $\O(1)$:
\equ{\bar{F}^{USp}_{S^3×Σ_{\fg}}=(\fg-1)\frac{4\sqrt{2}πQ^2p}{\sqrt{2(p+2)-f}}(1-\fn_f\tnu_f)\sqrt{1-\tnu_f^2}\,N^{\frac{3}{2}}\,.
}
\item When $0≤n<2$ with $f,p$ both being $\O(1)$:
\equ{\bar{F}^{USp}_{S^3×Σ_{\fg}}=(\fg-1)\frac{πQp(2-n\fn_f\tnu_f)}{\sqrt{3(2-n)}}\sqrt{4(4+n)Q^2 -2(2-n)}\,N\,.
}
Once again, the instanton contributions at $\O(N)$ will modify this result. 
\end{enumerate}
The above expressions can be compared to the appropriate $\bar{F}_{S^5_{\vec{ω}}}^{USp}$ expressions by using the same values for $ω_i$'s as in the case of the $SU$ theories.


\section{\texorpdfstring{$\bm{S^3_b}$ Free Energy}{S³b Free Energy}}\label{sec:S3pf}
Now we consider the 3d theories constructed in \cite{Sacchi:2021wvg} and compute their 3-sphere free energy at large $N$. The relevant ingredients can be found in \cite{Jain:2019lqb} and references therein. We will discuss these in proper detail in a companion paper \cite{Jain:2022td} dealing with 3d partition functions so this section will act as just a warm-up for that. For now, we just recall that the $S^3_b$ partition function can be written as follows:
\equ{Z_{S^3_b} = απρ(ιλ)
\label{ZS3bgen}}
Let us now see how the 3d free energy computations stack up against the computations of previous sections.

\subsection[\texorpdfstring{$SU(N)_k+N_f$}{SU(N)k+Nf}]{$\bm{SU(N)_k+N_f}$}
The 3d theory conjectured for this case with $k=1$ and $N_f=2N+2$ is given by the quiver diagram shown in Figure \ref{fig1}.
\fig{!h}{\includegraphics[scale=0.55]{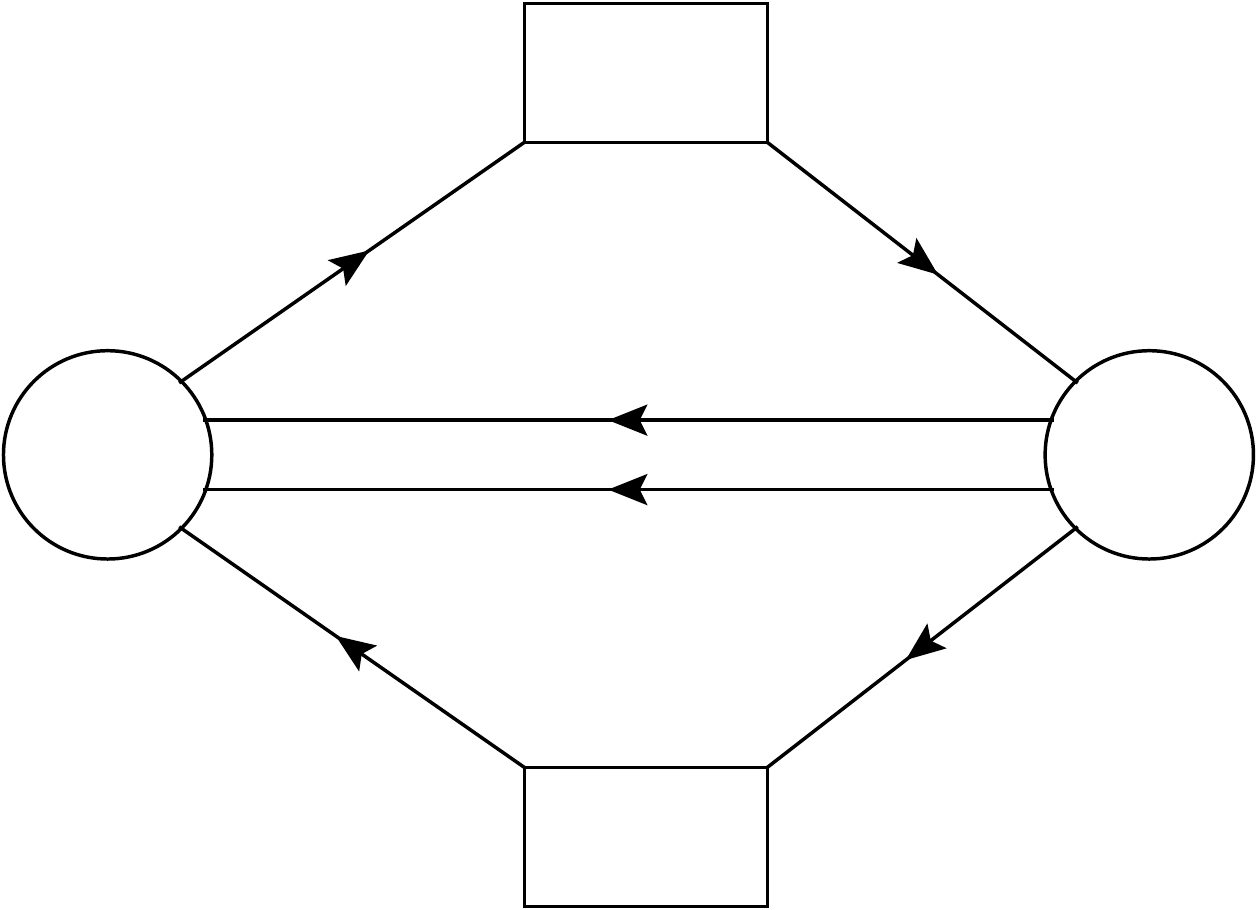}
\put(-115,129){$2N+1$}\put(-188,68){$N$}\put(-23,68){$N$}\put(-100,8){$1$}
\caption{The 3d theory associated with the compactification of 5d SCFT that UV completes the $SU(N)_1+(2N+2)F$ gauge theory on a torus with flux $(1,-1,0,⋯,0)$.}
\label{fig1}}

The free energy follows from \eqref{ZS3bgen}:
\equ{F_{S^3_b} = φooλ\s!
}

\section*{\centering Acknowledgements}
More short notes will be forthcoming monthly till the end of this year, regardless of any kind of scooping (hopefully). So, Like, Comment, and Subscribe to my Author Feed!


\references{5dRefs}

\end{document}